\documentclass[hidelinks,11pt,a4paper,oneside]{article}

\pdfoutput=1
\usepackage{amssymb}
\usepackage{amsmath}
\usepackage{cite}
\usepackage{graphicx}
\usepackage{braket}
\usepackage{amsbsy}
\usepackage{bm} 
\usepackage{euscript} 
\usepackage[a4paper,top=1in,bottom=1.25in,left=1in,right=1in]{geometry}
\usepackage[perpage,para]{footmisc}
\usepackage{lipsum}
\usepackage[margin=0.5in,font=small,labelfont=bf]{caption}
\usepackage{enumitem}

\newcommand\blfootnote[1]{%
  \begingroup
  \renewcommand\thefootnote{}\footnote{#1}%
  \addtocounter{footnote}{-1}%
  \endgroup
}

\begin{document}

\title{\textbf{Minimum energy path calculations\\with Gaussian process regression}}

\author{O-P.~Koistinen $^1$ \and E.~Maras $^2$ \and A.~Vehtari $^1$ \and H.~J\'onsson $^{2,3}$}

\date{}

\maketitle

\begin{center}
$^1$ Helsinki Institute for Information Technology HIIT,\\Department of Computer Science, Aalto University, Finland
\end{center}
\begin{center}
$^2$ Department of Applied Physics, Aalto University, Finland
\end{center}
\begin{center}
$^3$ Faculty of Physical Sciences, University of Iceland, 107 Reykjav\'{\i}k, Iceland
\end{center}
\begin{center}
hj@hi.is
\end{center}

\blfootnote{The article has been published in \textit{Nanosystems: Physics, Chemistry, Mathematics}. This version
includes some small corrections, clarifications and typographical corrections to the original text. Sentences with
notable changes have been indicated with a footnote.}

\begin{abstract}
The calculation of minimum energy paths for transitions such as atomic and/or spin rearrangements is
an important task in many contexts and can often be used to determine the mechanism and rate of transitions.
An important challenge is to reduce the computational effort in such calculations, especially when {\it ab initio} or
electron density functional calculations are used to evaluate the energy since they can require large
computational effort.
Gaussian process regression is used here to reduce significantly the number of energy evaluations needed to
find minimum energy paths of atomic rearrangements.
By using results of previous calculations to construct an approximate energy surface
and then converge to the minimum energy path on that surface in each Gaussian process iteration,
the number of energy evaluations is reduced significantly as compared with regular nudged elastic band
calculations. For a test problem involving rearrangements of a heptamer island on a crystal surface,
the number of energy evaluations is reduced to less than a fifth.
The scaling of the computational effort with the number of degrees of freedom as well as
various possible further improvements to this approach are discussed.

\textbf{Keywords:} minimum energy path, machine learning, Gaussian process, transition mechanism, saddle point

\end{abstract}


\hfill

\section{Introduction}

The task of predicting the rate and identifying the mechanism of transitions involving some rearrangements of atoms in or on the surface of solids shows up in many different applications, for example diffusion, crystal growth, chemical catalysis, nanotechnology, etc. 
At a finite temperature, the thermal fluctuations in the dynamics of atoms
can lead to rearrangements from one stable configuration to another, but these are rare events on the time scale of atomic vibrations, so direct dynamics simulations cannot in most cases be used for these types of studies. The separation of time scales typically amounts to several orders of magnitude, and a direct simulation would take impossibly long time. Instead, algorithms based on statistical mechanics as well as classical dynamics and focusing on the relevant rare events need to be applied \cite{TST,Kramers40,Keck67}. 
Typical transitions involve not just one or a few atoms but rather a large number of atoms, so the challenge is also to deal with multiple degrees of freedom.  One way of looking at the problem is to characterise the motion of the system on a high-dimensional energy surface, where the number of degrees of freedom is easily more than a hundred. A key concept is the reaction coordinate, which is usually taken to be a minimum energy path (MEP) on the energy surface connecting one minimum to another. The rate of transitions in solids is usually evaluated within harmonic transition state theory, which is based on a quadratic expansion of the energy surface at the initial state minimum and at the highest maximum along the MEP, which is a first order saddle point on the energy surface \cite{HTST}. 
For given initial and final states, the task is to determine the MEP and identify the saddle point(s) as well as possible unknown, intermediate minima \cite{Jonsson11}.  
The discussion here has been in terms of rearrangements of atoms, but similar considerations apply to reorientations of
magnetic moments \cite{bessarab_12,bessarab_13a,bessarab_14,bessarab_14b}.

The nudged elastic band (NEB) method is commonly used to find MEPs for atomic rearrangements \cite{Jonsson11,NEB95,NEBleri}.
An analogous method, referred to as the geodesic NEB, has been developed for magnetic transitions \cite{Bessarab15}.
In NEB calculations, some initial path is constructed between two local minima on the energy surface and the path is represented by a 
discrete set of replicas of the system. 
The replicas are referred to as images of the system. They consist of some set of values for all degrees of freedom in the system. 
The NEB algorithm then optimises iteratively the location of the images 
that are between the endpoint minima so as to obtain a discrete representation of the MEP.
Initially, the method was mainly used in combination with analytical potential energy functions, but today the method is used extensively 
in combination with electronic structure calculations.  A large amount of computer time is used in these calculations.
Each calculation typically involves about one hundred evaluations of the energy and force (the negative gradient of the energy)
for each one of the images, and the path is typically represented by five to ten images.  
Since a typical electronic structure calculation takes on the order of tens of CPU minutes or more, these calculations can be 
heavy.  
Also, several different possible final states usually need to be tested and the NEB calculation therefore repeated.  
In light of the widespread use of NEB and the large amount of CPU time used
in NEB calculations, it is of great practical importance to find ways to accelerate the calculations.  
The goal should be to use the information coming from all the computationally intensive electronic structure calculations in an optimal way 
so as to reduce as much as possible the number of iterations needed to reach the MEP.  

It has recently been shown that a machine learning algorithm based on neural networks can be used to significantly reduce the
computational effort in NEB calculations \cite{Peterson16}.
An approximate representation of the energy surface is constructed from the
calculations using a machine learning approach, and the MEP is calculated using the NEB method on this approximate
surface. Then, additional evaluations are made of the true energy surface, the approximate model surface refined, etc., until 
convergence on the MEP of the true energy surface has been reached.
The number of function evaluations was shown to drop dramatically by applying such an approach \cite{Peterson16}.

We present here an initial step in the development of a similar approach to accelerated MEP calculations based on
Gaussian process (GP) regression \cite{OHagan78,MacKay98,Neal99,Rasmussen06}. 
This approach could have some advantages over neural networks for such applications. 
Neural networks have a large number of weights which can have multimodal distributions making the search for global optimum
difficult and leading to possible dependence on the initial values of the parameters \cite{Peterson16}.
Also, the handling of uncertainties in GP theory is easier than in neural networks since the prediction equations are analytical 
and integration over the parameter space can be carried out more easily.  It is, therefore, of interest to test the 
efficiency of the GP approach in MEP calculations. We report in this article initial feasability studies. 
More extensive testing and comparison with other approaches such as neural networks is left for future work. 

The article is organized in the following way: The methodology is presented in the next section, followed by a section on applications,
both a simple two-dimensional system and a larger test problem involving rearrangements of a heptamer island on a crystal surface.
The article concludes with a discussion section.


\section{Methods}

The method presented here for finding the minimum energy paths can be viewed as an acceleration of an 
NEB calculation by making use of Gaussian process theory.  
Previously calculated data points are used to construct an
approximate model of the energy surface, and the MEP is found for this approximate surface
before additional calculations of the true energy are carried out.
This gives an interpolation between the calculated points and also provides an extrapolation that can be used to explore the energy surface with larger moves. 
The savings in computational effort are based on
the fact that several computationally light iterations can be made for the approximate surface
in between the computationally demanding evaluations of the true energy function.
A brief review of the NEB method is first given, then a description of the Gaussian process regression, and finally a
detailed algorithm describing how the calculations were carried out in the present case. 


\subsection{Nudged elastic band method}
\label{sect:NEB}

Given two local minima on the energy surface, the task is to find an MEP connecting the two.
The definition of an MEP is that the gradient has zero component perpendicular to the path tangent at each point along the path.
The NEB method needs to be started with some initial path between the two minima that is represented by a set of images.
Most often, a straight line interpolation between the minima is used to generate the initial path \cite{NEBleri}, but 
a better approach is to start with a path that interpolates as closely as possible the distances between atoms \cite{Smidstrup14}.

The key aspect of the NEB algorithm is the nudging, a force projection which is used to decouple the 
displacements of the images perpendicular to the path towards the MEP
from the displacements that affect their distribution along the path. 
In order to make this projection, an estimate of the local tangent to the path at each of the images is needed.
A numerically stable choice involves finding the line segment from the current image to the adjacent image of higher energy \cite{ImprovedTangent}. 

Given this decoupling, there are
several different options for distributing the images along the path. Some constraint is needed to prevent the images from
sliding down to the minima at the two ends. In most cases an even distribution is chosen,
but one can also choose to have, for example, higher density of images where the energy is larger \cite{ClimbingImage}.
An attractive spring force is typically introduced between adjacent images to control the spacing between images, and
this also prevents the path from becoming arbitrarily long in regions of little or no force. 
The latter is important, for example, in calculations of adsorption and desorption of molecules at surfaces.
For systems that can freely translate and rotate, such as nano-clusters in free space, it is important to remove the translational and rotational degrees of freedom. This is non-trivial because the system cannot be treated as a rigid body. 
A method for doing this efficiently based on quaternions has recently been presented \cite{Melander15}. 

The component of the force acting on each image perpendicular to the path is used to iteratively move the images from the initial path
to the MEP. The force is the negative of the gradient, and in most cases an evaluation of the energy delivers also the gradient vector at
little or no extra expense. The largest amount of information from an evaluation of a point on the energy surface is, therefore,
represented by the gradient. It is, however, typically too expensive to evaluate second derivatives of the energy, and iterative algorithms for 
moving the images towards the MEP are therefore based solely on the gradient and the energy at each point.
A simple and numerically stable method that has been used extensively in NEB calculations will be used here.
It is based on a velocity Verlet method 
where only the component of the velocity in the direction of the force is included and the velocity is 
zeroed if its dot product with the force becomes negative \cite{NEBleri}. A somewhat higher
efficiency can be obtained by using a quadratically convergent algorithm such as conjugate gradients or quasi-Newton \cite{Sheppard08},
but those can be less stable especially in the beginning of an NEB calculation.
A linear interpolation between the initial state minima was used in all the calculations presented here, and the number of images, $N_{\mathrm{p}}$,
was chosen to be either seven or ten. An equal distribution of the images along the path was chosen.

The focus here is on calculations where the energy and the gradient are obtained using some {\it ab initio} or density functional theory 
calculation. The computational effort in all other parts of the calculation is then insignificant in comparison, and thus the overall computational
effort is well characterised by simply the number of times the energy and force need to be evaluated in order to converge on the MEP. 
Below, we introduce a strategy to accelerate the MEP search with Gaussian process regression.


\subsection{Gaussian processes regression}
\label{sect:GPtheory}

The general idea behind the strategy used here is similar to the one introduced by Peterson \cite{Peterson16}.
The idea is to use the calculations carried out so far to train an approximate model of the energy surface, and
find the MEP with the conventional methods using the approximations of the energy and gradient based on this model.
After converging to the MEP on the approximate energy surface, the true energy and force are evaluated again
showing whether or not the path has converged to the true MEP.
If not, the model is updated with the new values of the true energy and force to get a more accurate approximation, and this
is continued iteratively until the true MEP has been found.
Since the number of true energy and force evaluations is the measure of computational effort, basically any method
can be used to optimise the path on the approximate energy surface, as long as it converges to an MEP.

Here, a Gaussian process is used as a probabilistic model for the energy surface.
GPs provide a flexible framework for modelling multidimensional functions.
Through the selection of the covariance function and its hyperparameters, smoothness properties of the function can easily be defined, and those properties can also be learned from the data.
It is also straightforward to both include derivative observations
into the model and to predict derivatives of the modelled function.
Analytical expressions for the posterior predictions conditional on the hyperparameters allow
both fast predictions and reliable estimation of uncertainties.
In cases where only a small number of observations are available,
Gaussian processes have been shown to have good predictive performance compared to
other machine learning methods \cite{Lampinen01}.

A Gaussian process can be seen as a probability distribution
over functions in a continuous domain (see, e.g., \cite{OHagan78,MacKay98,Neal99,Rasmussen06}).
In a GP, the joint probability distribution of the function values
$f(\mathbf{x}^{(1)}), f(\mathbf{x}^{(2)}), \dots , f(\mathbf{x}^{(N)})$ at any finite set of input points
$\mathbf{x}^{(1)}, \mathbf{x}^{(2)}, \dots , \mathbf{x}^{(N)} \in \mathbb{R}^D$ is a multivariate Gaussian distribution.
A GP is defined by a mean function $m(\mathbf{x})$ and a covariance function
$k(\mathbf{x}^{(i)},\mathbf{x}^{(j)})$, which determines the covariance between $f(\mathbf{x}^{(i)})$ and $f(\mathbf{x}^{(j)})$,
e.g., based on the distance between $\mathbf{x}^{(i)}$ and $\mathbf{x}^{(j)}$.

Consider a regression problem $y = f(\mathbf{x}) + \epsilon$, where $\epsilon$ is Gaussian noise with variance $\sigma^2$,
and a training data set $\{\mathbf{X},\mathbf{y}\}$, where $\mathbf{X} \in \mathbb{R}^{N \times D}$ denotes a matrix of $N$ input vectors
$\mathbf{x}^{(1)}, \mathbf{x}^{(2)}, \dots , \mathbf{x}^{(N)} \in \mathbb{R}^D$ and $\mathbf{y}$ is a vector of the
corresponding $N$ noisy observations.
By choosing a Gaussian process to model function $f$, different prior assumptions can be made about the properties of the function,
and after observing $\{\mathbf{X},\mathbf{y}\}$, the posterior predictive probabilities for the function values at a set of new points
can be calculated analytically as a multivariate Gaussian distribution.
Here, the mean function is taken to be $m(\mathbf{x}) = 0$ and the covariance function is assumed to have the form
\begin{equation*}
k(\mathbf{x}^{(i)},\mathbf{x}^{(j)}) = c^2 + \eta^2\exp\left(-\frac{1}{2}\sum_{d=1}^{D} \rho_d^{-2}(x_d^{(i)}-x_d^{(j)})^2\right),
\end{equation*}
where $\eta^2$ and $\bm{\rho} = \{ \rho_1, \ldots, \rho_D \}$ are 
the hyperparameters of the GP model.
The squared exponential covariance function is infinitely differentiable and thus favours smooth functions.
The length scales $\bm{\rho}$ define how fast the function $f$ can change, and $\eta^2$ controls the magnitude of the overall variation.
The additional constant term $c^2$ has a similar effect as integration over an unknown constant mean function having a Gaussian 
prior distribution with variance $c^2$.
The posterior predictive distribution for a value of the function at a new point $\mathbf{x}^*$, denoted as $f^*$,
is described by a Gaussian distribution with mean
\begin{equation*}
  {\mathrm{E}}[f^*|\mathbf{x}^*,\mathbf{y},\mathbf{X},\bm{\theta}] = K(\mathbf{x}^*,\mathbf{X})(K(\mathbf{X},\mathbf{X}) + \sigma^2 \mathbf{I})^{-1}\mathbf{y}
\end{equation*}
and variance
\begin{equation*}
\label{gpvar}
  \mathrm{Var}[f^*|\mathbf{x}^*,\mathbf{y},\mathbf{X},\bm{\theta}] =
  k(\mathbf{x}^*,\mathbf{x}^*)-K(\mathbf{x}^*,\mathbf{X}) (K(\mathbf{X},\mathbf{X}) + \sigma^2 \mathbf{I})^{-1}K(\mathbf{X},\mathbf{x}^*),
\end{equation*}
where $\mathbf{I}$ is the identity matrix and the notation $K(\mathbf{X},\mathbf{X}')$ represents a covariance matrix
with entries $K_{ij} = k(\mathbf{x}^{(i)},\mathbf{x}'^{(j)})$. The hyperparameter values $\bm{\theta} = \{\eta^2,\bm{\rho}\}$ are optimised by defining a
prior probability distribution $p(\bm{\theta})$ and maximising the marginal posterior probability density
$p(\bm{\theta}|\mathbf{y},\mathbf{X}) = p(\bm{\theta})p(\mathbf{y}|\mathbf{X},\bm{\theta})$ after observing $\mathbf{y}$.

Since differentiation is a linear operation, the derivative of a Gaussian process is also a Gaussian process 
(see, e.g., \cite{Rasmussen03,Solak03}),
and this makes it possible to use observations of the derivative of the function and also to predict derivatives of the function $f$.
The partial derivative observations can simply be included in the observation vector $\mathbf{y}$ and the
covariance matrix correspondingly extended with the covariances between the observations and the partial derivatives and the 
covariances between the partial derivatives themselves. In the case of the squared exponential covariance function, 
these entries are obtained by

\begin{align*}
\mathrm{Cov}\left[\frac{\partial f^{(i)}}{\partial
  x^{(i)}_d},f^{(j)}\right]=&\frac{\partial}{\partial
  x^{(i)}_d}\mathrm{Cov}\left[f^{(i)},f^{(j)}\right]
  = \frac{\partial}{\partial x^{(i)}_d} k(\mathbf{x}^{(i)},\mathbf{x}^{(j)}) \\
  =& \eta^2\exp\left(-\frac{1}{2}\sum_{g=1}^{D}\rho_g^{-2}(x_g^{(i)}-x_g^{(j)})^2\right)(-\rho_d^{-2}(x_d^{(i)}-x_d^{(j)})),
\end{align*}
and
\begin{align*}
\mathrm{Cov}\left[\frac{\partial f^{(i)}}{\partial
  x^{(i)}_{d_1}},\frac{\partial f^{(j)}}{\partial x^{(j)}_{d_2}}\right]=&
  \frac{\partial^2}{\partial x^{(i)}_{d_1}\partial x^{(j)}_{d_2}}\mathrm{Cov}\left[f^{(i)},f^{(j)}\right] =
  \frac{\partial^2}{\partial x^{(i)}_{d_1}\partial x^{(j)}_{d_2}} k(\mathbf{x}^{(i)},\mathbf{x}^{(j)}) \\
  =& \eta^2\exp\left(-\frac{1}{2}\sum_{g=1}^{D}\rho_g^{-2}(x_g^{(i)}-x_g^{(j)})^2\right)\times\\
  &\rho_{d_1}^{-2}\left(\delta_{{d_1}{d_2}}-\rho_{d_2}^{-2}(x^{(i)}_{d_1}-x^{(j)}_{d_1})(x^{(i)}_{d_2}-x^{(j)}_{d_2})\right),
\end{align*}
where $\delta_{{d_1}{d_2}}=1$ if $d_1=d_2$, and $\delta_{{d_1}{d_2}}=0$ if $d_1 \neq d_2$.

These same expressions are useful also when predicting values of the derivatives.
The posterior predictive distribution of the partial derivative of function $f$ with respect to dimension $d$
at a new point $\mathbf{x^*}$ is a Gaussian distribution with mean
\begin{equation*}
  {\mathrm{E}}\left[ \frac{\partial f^*}{\partial x^*_d} \middle| \mathbf{x}^*,\mathbf{y},\mathbf{X},\bm{\theta} \right] =
  \frac{\partial K(\mathbf{x}^*,\mathbf{X})}{\partial x^*_d}(K(\mathbf{X},\mathbf{X}) + \sigma^2 \mathbf{I})^{-1}\mathbf{y}
\end{equation*}
and variance
\begin{equation*}
  \mathrm{Var}\left[\frac{\partial f^*}{\partial x^*_d} \middle|\mathbf{x}^*,\mathbf{y},\mathbf{X},\bm{\theta}\right] =
  \frac{\partial^2 k(\mathbf{x}^*,\mathbf{x}^*)}{\partial x^*_d \partial x^*_d} -
  \frac{\partial K(\mathbf{x}^*,\mathbf{X})}{\partial x^*_d} (K(\mathbf{X},\mathbf{X}) + \sigma^2 \mathbf{I})^{-1}\frac{\partial K(\mathbf{X}, \mathbf{x}^*)}{\partial x^*_d}.
\end{equation*}

In the present application, the vector $\mathbf{x}$ represents coordinates of the atoms and the function $f$ the energy of the system.
The extended version of observation vector $\mathbf{y}$ includes the true values of the energy 
as well as the partial derivatives of the energy with respect to the coordinates of the atoms
at the various sets of 
coordinates $\mathbf{x}^{(1)}, \mathbf{x}^{(2)}, \dots , \mathbf{x}^{(N)}$.
With this input, the Gaussian process model is used to predict the most likely value of energy $f^*$ and its derivatives
$\frac{\partial f^*}{\partial x^*_d}$ at a new set of atom coordinates $\mathbf{x}^*$ representing in this case an image
in the discrete path representation between the initial and final state minima.
Since the training data are assumed to be noiseless and include also derivative observations, the equations for the mean predictions
can be presented as
\begin{equation}\label{gpmean}
  {\mathrm{E}}[f^*|\mathbf{x}^*,\mathbf{y}_{\mathrm{ext}},\mathbf{X},\bm{\theta}] = {\mathbf{K}}_{\mathrm{ext}}^* {\mathbf{K}}_{\mathrm{ext}}^{-1} \mathbf{y}_{\mathrm{ext}}
\end{equation}
and
\begin{equation}\label{gradmean}
  {\mathrm{E}}\left[ \frac{\partial f^*}{\partial x^*_d} \middle| \mathbf{x}^*,\mathbf{y}_{\mathrm{ext}},\mathbf{X},\bm{\theta} \right] =
  \frac{\partial {\mathbf{K}}_{\mathrm{ext}}^*}{\partial x^*_d} {\mathbf{K}}_{\mathrm{ext}}^{-1} \mathbf{y}_{\mathrm{ext}},
\end{equation}
where
\begin{equation*}
\mathbf{y}_{\mathrm{ext}} = \begin{bmatrix}
            y^{(1)} \cdots y^{(N)},
            \frac{\partial f^{(1)}}{\partial x^{(1)}_1}  \cdots \frac{\partial f^{(N)}}{\partial x^{(N)}_1},
            \frac{\partial f^{(1)}}{\partial x^{(1)}_2}  \cdots \frac{\partial f^{(N)}}{\partial x^{(N)}_2}, & \cdots&,
            \frac{\partial f^{(1)}}{\partial x^{(1)}_D}  \cdots \frac{\partial f^{(N)}}{\partial x^{(N)}_D}
            \end{bmatrix}^{\mathrm{T}},
\end{equation*}
\begin{equation*}
{\mathbf{K}}_{\mathrm{ext}}^* = \begin{bmatrix}
            K(\mathbf{x}^*,\mathbf{X}) & \frac{\partial K(\mathbf{x}^*,\mathbf{X})}{\partial x_1} & \frac{\partial K(\mathbf{x}^*,\mathbf{X})}{\partial x_2} & \cdots & \frac{\partial K(\mathbf{x}^*,\mathbf{X})}{\partial x_D}
            \end{bmatrix},
\end{equation*}
and
\begin{equation*}
{\mathbf{K}}_{\mathrm{ext}} = \begin{bmatrix}
            K(\mathbf{X},\mathbf{X}) & \frac{\partial K(\mathbf{X},\mathbf{X}')}{\partial x_1'} & \frac{\partial K(\mathbf{X},\mathbf{X}')}{\partial x_2'} & \dots & \frac{\partial K(\mathbf{X},\mathbf{X}')}{\partial x_D'} \\[0.3em]
            \frac{\partial K(\mathbf{X},\mathbf{X}')}{\partial x_1} & \frac{\partial^2 K(\mathbf{X},\mathbf{X}')}{\partial x_1 \partial x_1'}    & \frac{\partial^2 K(\mathbf{X},\mathbf{X}')}{\partial x_1 \partial x_2'}  & \cdots & \frac{\partial^2 K(\mathbf{X},\mathbf{X}')}{\partial x_1 \partial x_D'} \\[0.3em]
            \frac{\partial K(\mathbf{X},\mathbf{X}')}{\partial x_2} & \frac{\partial^2 K(\mathbf{X},\mathbf{X}')}{\partial x_2 \partial x_1'}    & \frac{\partial^2 K(\mathbf{X},\mathbf{X}')}{\partial x_2 \partial x_2'}  & \cdots & \frac{\partial^2 K(\mathbf{X},\mathbf{X}')}{\partial x_2 \partial x_D'} \\[0.3em]
            \vdots                                                  & \vdots                                                            & \vdots                                                          & \ddots & \vdots \\[0.3em]
            \frac{\partial K(\mathbf{X},\mathbf{X}')}{\partial x_D} & \frac{\partial^2 K(\mathbf{X},\mathbf{X}')}{\partial x_D \partial x_1'}    & \frac{\partial^2 K(\mathbf{X},\mathbf{X}')}{\partial x_D \partial x_2' } & \cdots & \frac{\partial^2 K(\mathbf{X},\mathbf{X}')}{\partial x_D \partial x_D'}
          \end{bmatrix}.
\end{equation*}


\vskip 0.6 true cm

\subsection{Algorithm for GP-aided MEP search}
\label{sect:Algorithm}

\begin{list}{}{\topsep=0in \leftmargin=0.2in \rightmargin=0.2in}

\vskip 0.3 true cm

\item {\bf Input:} the coordinates, energy and gradient at the two minima on the energy surface, the number of images representing the path ($N_{\mathrm{p}}$), a convergence limit ($CL$).
\vskip 0.2 true cm

\item {\bf Output:} a minimum energy path represented by $N_{\mathrm{p}}$ images.
\vskip 0.2 true cm

\end{list}

\begin{enumerate}[topsep=0in, leftmargin=0.2in, rightmargin=0.2in]

\item Place the initial $N_{\mathrm{p}}$ images equally spaced along a straight line between the two minima.

\vskip 0.2 true cm

\item Repeat until convergence (outer iteration loop):
  \begin{enumerate}[label={\Alph*.}]
  \item Evaluate the true energy and its gradient at the $N_{\mathrm{p}} - 2$ intermediate images of the path, and add them to the training data.
  \item Calculate the negative energy gradient (i.e., force) component perpendicular to the path ($ngc$) for each intermediate image,
         and denote the mean of their norms as $M_{ngc}$.
  \item If $M_{ngc} < CL$, the path has converged to the true MEP.
  \item Optimise the hyperparameters of the GP model based on the training data, and calculate the matrix inversion in equation \ref{gpmean}.
  \item Define $CL_{\mathrm{relax}}$ as $\frac{1}{10}$ of the smallest
           $M_{ngc}$ so far, and repeat (relaxation phase):
    \begin{enumerate}[label={\Roman*.}]
    \item Move the intermediate images according to any stable path optimisation algorithm.
    \item Update the GP posterior mean energy and gradient at the new intermediate images using equations \ref{gpmean} and \ref{gradmean}.
    \item Calculate $ngc$ for each image
             using the GP posterior mean gradient,
             and denote the mean of their norms as $M_{ngc}^{\mathrm{GP}}$.
    \item If $M_{ngc}^{\mathrm{GP}} < CL_{\mathrm{relax}}$, or if $M_{ngc}^{\mathrm{GP}}$ is increasing,
              exit the relaxation phase (E).
    \end{enumerate}
  \end{enumerate}
\end{enumerate}

\vskip 0.2 true cm
The GP calculations make use of the GPstuff toolbox \cite{Vanhatalo13}.
For the hyperparameter optimisation which is carried out after each evaluation of the true energy and force, 
the computational effort scales as $\mathcal{O}((N(D+1))^3)$,
where $N$ is the number of observations and $D$ is the number of degrees of freedom (here coordinates of movable atoms).
Since the hyperparameters and observations stay the same during a search for the MEP on the approximate energy surface,
the matrix inversion in equation \ref{gpmean} needs to be computed only once for each such optimisation of the path.
Thus, the complexity of one inner iteration on the GP posterior energy surface is $\mathcal{O}(N(D+1))$. 

The length of any one displacement of an image is restricted to be less than half of the initial interval between 
the images in order to prevent the path from forming loops.
Convergence of the path to the MEP is determined from the norm of the force component perpendicular to the
path at each of the intermediate images. 
The path is considered to be converged to the MEP, when the mean of the true values of these norms is less than 0.001 eV/{\AA}.
During the relaxations, norms based on the current GP model are monitored and the mean of these used as a convergence measure.
Since it is not necessary to find a path that is accurately converged on the MEP of the inaccurate, approximate energy surface,
the convergence limit for each relaxation phase is defined as $\frac{1}{10}$ of the smallest true mean of norms evaluated so far.
Higher convergence limits at early relaxation steps speed up the algorithm and they also make it more stable by preventing
the path from escaping too far from the true observation points.
For the same reason, the relaxation is stopped before convergence if the convergence measure starts to increase.


\section{Applications}
\label{sect:application}

The method described above has been applied to two test problems: A simple two-dimensional problem where the energy surface
can be visualised, and a more realistic problem involving the rearrangements of atoms in a heptamer island on a crystal 
surface.

\subsection{Two-dimensional test problem}

The two-dimensional problem is formulated by coupling a degree of freedom representing the simultaneous formation and breaking of  
chemical bonds
with a degree of freedom representing a harmonic oscillator solvent environment.  
The model along with the detailed equations is described in the appendix A.2 of reference \cite{NEBleri}.  Here, one additional repulsive
Gaussian was added to shift the saddle point away from the straight line interpolation between the two minima. A contour graph
of the energy surface is shown in figure 1. 

\begin{figure}[htbp]
\centering
\includegraphics[width=0.95 \columnwidth]{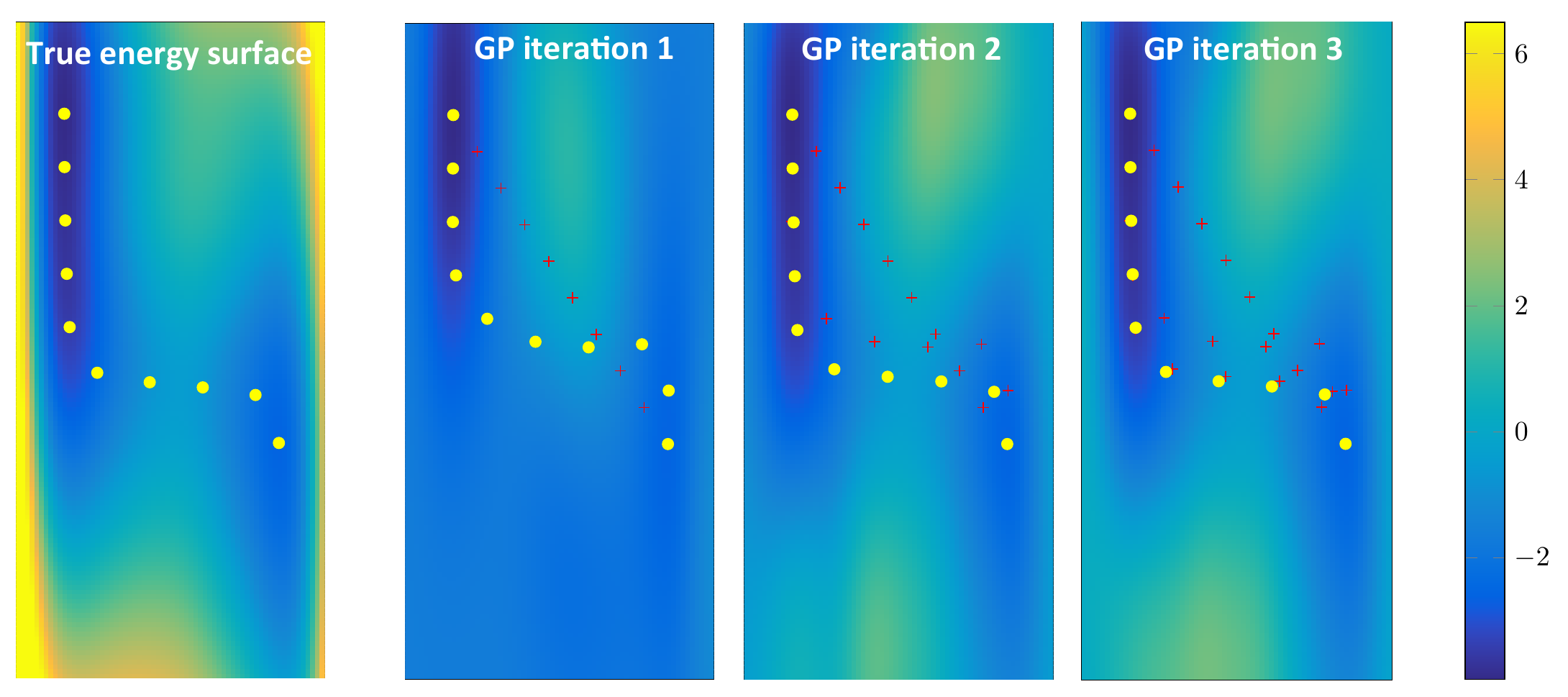}
\caption{
The true and Gaussian process approximated energy surface and minimum energy path for a two-dimensional test problem.
Far left:  
The true energy surface and points on the minimum energy path (yellow dots).
Far right and intermediate figures: 
The approximate energy surface generated by the Gaussian process regression after one, two and three iterations, points ('images')
on the estimated minimum energy path and points where the true energy and force have been 
calculated (red + signs) at each stage of the calculations.
}
\label{fig:fig1}
\end{figure}

This example shows how the GP model of the energy surface is gradually built up and refined as more observations, 
i.e., calculations of the true energy and partial derivatives of the energy, are made.
Here, $N_{\mathrm{p}}=10$ images are used to represent the path, and the calculation is started by placing the images along a
straight line between the two minima on the energy surface. The first observations are made at those points 
(see red + signs on the figure second from the left). Based on the
energy and partial derivatives of the energy at those points, the GP model already shows some of the most important features of
the energy surface close to the linear interpolation, but completely misses the increase in energy in the lower half of the figure.
The relaxation of the images on this rough estimate of the energy surface does not, however, bring the images far from the initial
placement because of the condition that the relaxation phase is stopped early if the convergence measure, i.e.,
the mean of the magnitudes of the force components perpendicular to the path at the intermediate images,
increases.$^\dagger$ \blfootnote{$^\dagger$ A corrected sentence.}
In the second GP iteration, observations are made at the position of the images at the end of 
the first GP iteration. When those data points are fed into the GP model, the energy surface is already showing the essential features 
around the MEP, but of course misses the steep increase in the energy far from the MEP.  
The relaxation of the images during the second GP
iteration brings them quite close to the MEP. The addition of observations at those points 
at the beginning of the third GP iteration refines the model energy surface further. 
While a total of six GP iterations is required to bring the images onto the MEP within the tight tolerance of 0.001 eV/{\AA} in the 
mean magnitude of the force components perpendicular to the path, no visible changes occur in the contour graph
or the location of the images, so the results are not displayed in the figure. 


\subsection{Heptamer island on a crystal surface}

A more realistic test problem, which has been used in several studies of MEP and saddle point searches, involves an
island of seven atoms on the (111) surface of a face-centered cubic (FCC) crystal (see, for example, references \cite{Henkelman00,Chill2014}).
Roughly, this represents a metallic system, but the interaction between the atoms is described here with a simple Morse potential
to make it easier to implement the benchmark calculation.
The initial, saddle point and final configurations of the atoms for three possible rearrangements of the atoms are shown in figure 2.
Several other transitions are possible (see reference \cite{Henkelman00}), but these three are chosen as examples.

\begin{figure}[htbp]
\centering
\includegraphics[width=0.95 \columnwidth]{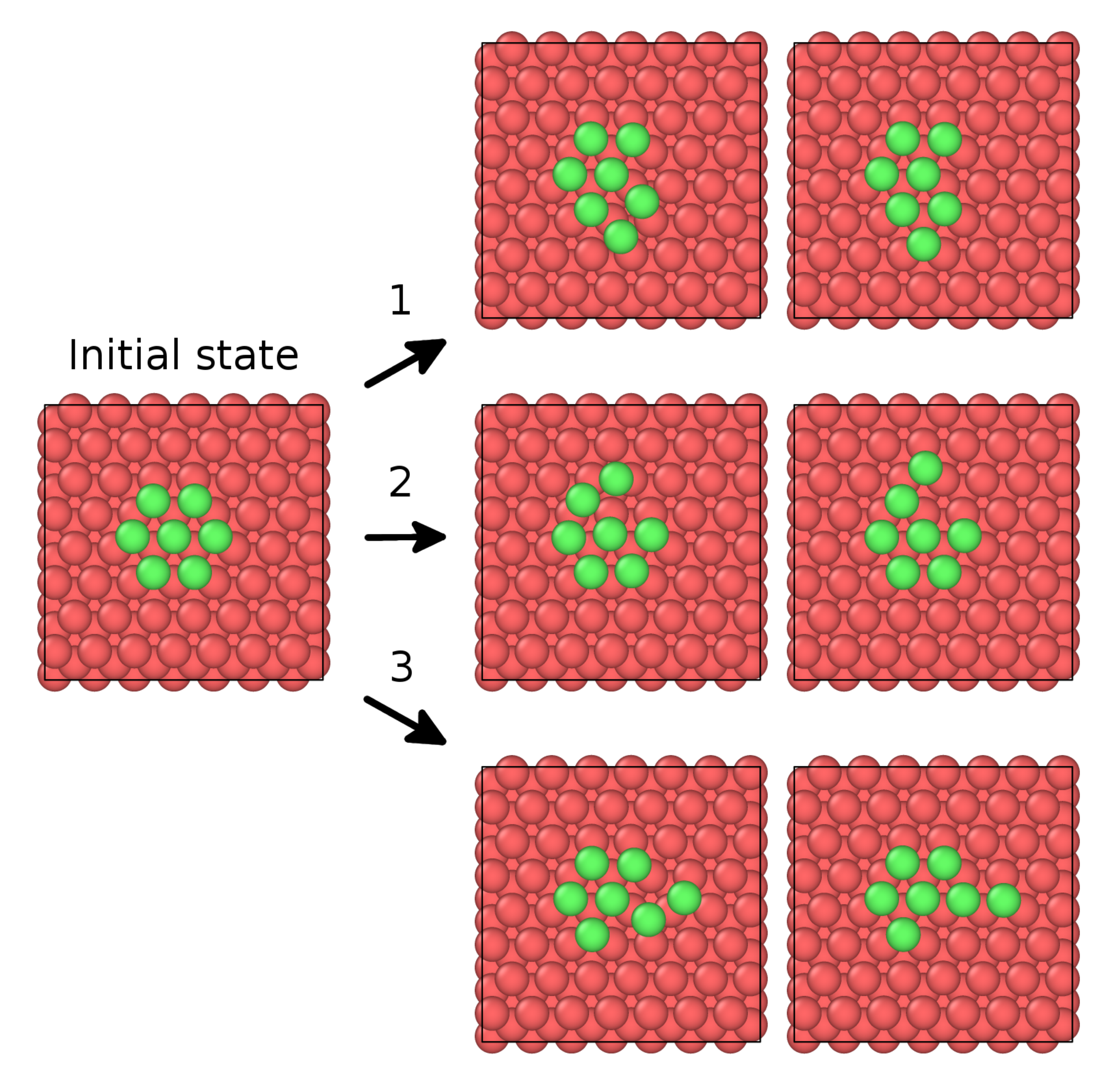}
\caption{
On-top view of the surface and the seven-atom island used to test the efficiency of the Gaussian process regression method. 
The initial state is shown to the left. The saddle point configurations and the final state configurations of three example transitions 
are also shown. Transition 1 corresponds to a pair of edge atoms sliding to adjacent FCC sites. In transition 2, an atom 
half way dissociates from the island. In transition 3, a pair of edge atoms moves in such a way that one of the atoms is displaced away from the island while the other atom takes its place. At the same time the other island atoms as well as some of the underlying atoms
also move but in the end return to nearly the same position as they had initially.
}
\label{fig:fig2}
\end{figure}

The three examples chosen here represent three types of transitions that can occur in the shape of the island.
In one case, a pair of edge atoms slides to adjacent FCC sites, in another an atom half way dissociates from the island, and in the 
third case a pair of edge atoms moves in such a way that one of the atoms is displaced away from the island while the other atom takes its place.  

The energy along the MEP for transition 3 is shown in figure 3 as well as the energy of the $N_{\mathrm{p}}=7$ images at the end of GP 
iterations 1 to 7. After the first and second GP iteration, the estimates of the MEP are quite inaccurate and the energy rises along those
paths by more than 3 eV, but already after the third GP iteration, the estimated energy barrier is not too far from the accurate 
value.
After the fifth GP iteration, the shape of the energy curve is quite well reproduced, and after seven iterations the energy along the MEP
of the approximate energy surface is nearly indistinguishable from the energy along the true MEP.

\begin{figure}[htbp]
\centering
\includegraphics[width=0.65 \columnwidth]{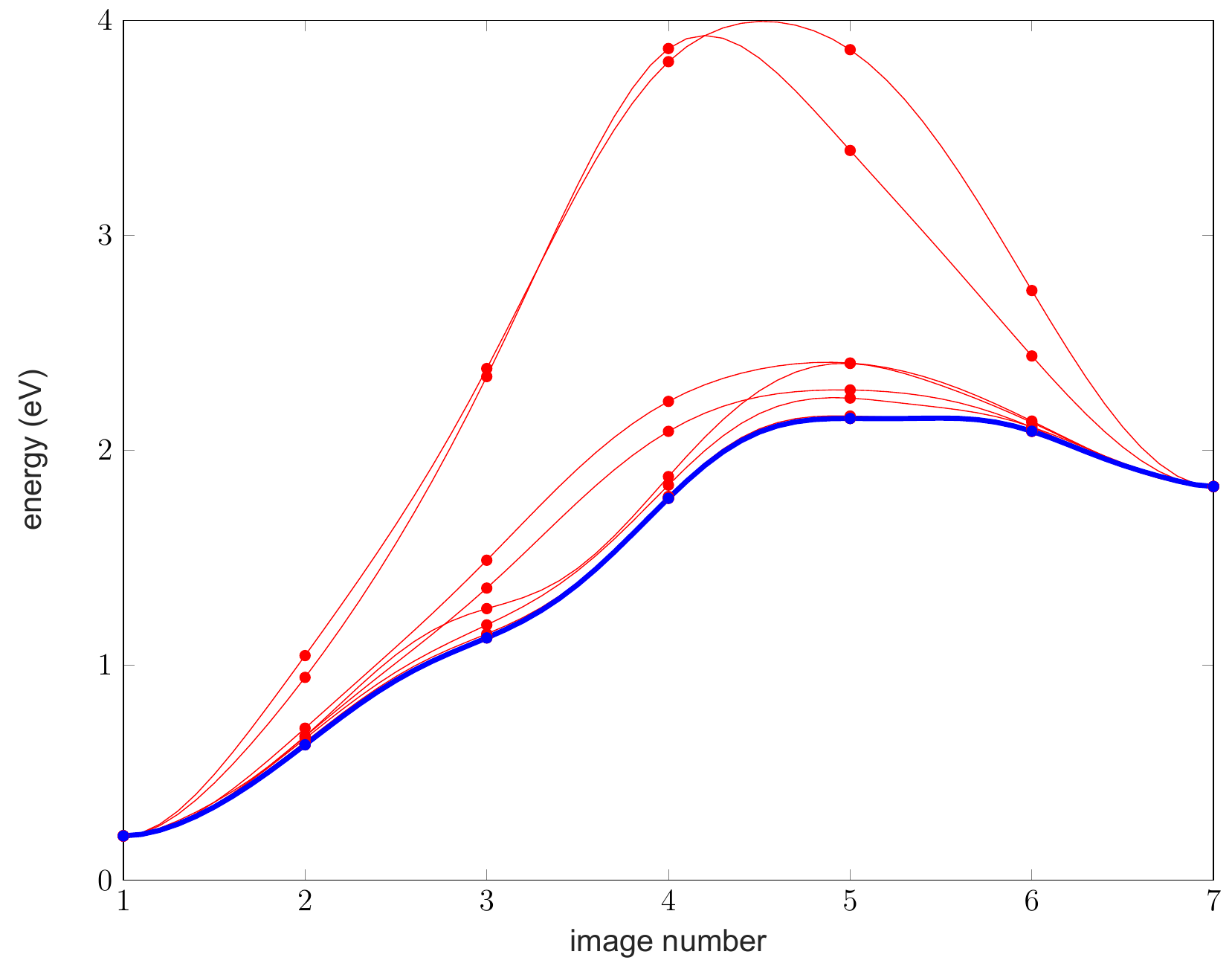}
\caption{
Energy along paths for transition 3 shown in figure 2. The energy of images on the true MEP are shown in blue, but
the energy of images on MEPs of approximate models of the energy surface obtained after one to seven Gaussian process iterations 
are shown in red. After the first two 
Gaussian process iterations, the energy barrier for this transition is greatly overestimated, but already after three iterations the
estimated energy barrier is quite close to the true value, and after seven iterations an accurate estimate is obtained from the model 
energy surface.
}
\label{fig:fig3}
\end{figure}

The number of energy and force evaluations needed to converge the five intermediate 
images to the MEP in both a regular NEB calculation and in a 
GP-aided calculation was found for varying number of degrees of freedom. The average of the three transitions depicted
in figure 2 is shown in figure 4. The number of degrees of freedom varies from 21 (as only the island atoms are allowed to move
while all the substrate atoms are kept immobile) to 42 (as seven of the closest substrate atoms are also allowed to move during the
transition). The number of energy and force evaluations for the NEB method obtained here is similar to what has been reported
earlier for this test problem, see references \cite{Henkelman00,Chill2014}. It is possible to use a more efficient minimisation scheme
to relax the images in NEB calculations \cite{Sheppard08}, but the difference is not large.

A large reduction in the number of energy and force evaluations is obtained by using the GP regression, as shown in figure 4. 
With the GP regression, the reduction is to less than a fifth as compared with the regular NEB calculation. 
In calculations involving {\it ab initio} or density functional theory evaluation of the energy and force, the computational 
effort is essentially proportional to this number of observations and the additional calculations involved in the GP regression are 
insignificant in comparison. This test problem, therefore, shows that the use of GP regression can significantly reduce the 
computational effort in, for example, calculations of surface processes.

\begin{figure}[htbp]
\begin{minipage}{\linewidth}
\centering
\includegraphics[width=0.65 \columnwidth]{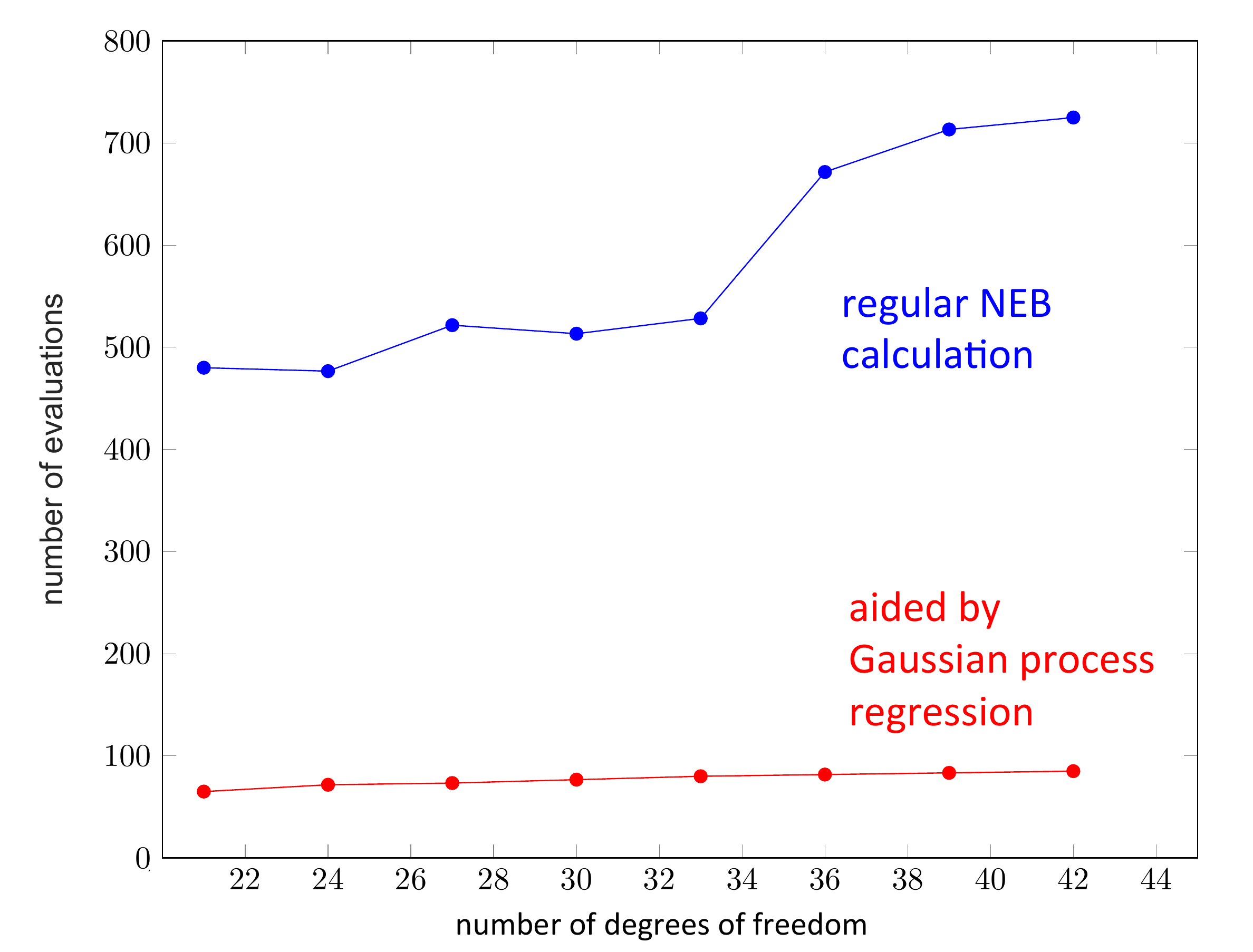}
\caption{
The average number of energy and force evaluations needed to converge five intermediate images on the minimum energy paths of the three
heptamer island transitions shown in figure 2
as a function of the number of degrees of freedom included in the calculations. 
The convergence tolerance is 0.001 eV/{\AA} for the mean of the magnitudes of
the perpendicular force components at the intermediate images.$^\dagger$
For the smallest number, 21, only the seven island atoms are allowed to move and all substrate atoms are immobile.
For a larger number of degrees of freedom, some of the substrate atoms are also allowed to move during the transition.
In the regular NEB calculations (blue dots), the minimization method for relaxing the images to the MEP is
based on a velocity Verlet algorithm, as described in reference \cite{NEBleri}.  
In the Gaussian process regression calculations (red dots), the number of true energy and function evaluations is less than a fifth of
what is needed in the regular NEB calculation. This illustrates well the large reduction in the computational effort that Gaussian process
regression can provide in a typical surface process calculation.
}
\label{fig:fig4}
\footnotetext[0]{$^\dagger$ A corrected sentence.}
\end{minipage}
\end{figure}


\section{Discussion}
\label{sec:disc}

The results presented in this article indicate that GP regression is a powerful approach for significantly reducing the 
computational effort in calculations of MEPs for transitions. This is important since a great deal of computer time is 
used in such calculations, especially when {\it ab initio} or density functional theory calculations are used to evaluate
the energy and atomic forces. The heptamer island test problem studied here indicates that the computational effort can
be reduced to less than a fifth. But, this study represents only an initial proof-of-principle demonstration of the GP regression in 
this context. There are several ways in which the implementation can be improved and made more efficient. One of the 
advantages of GP regression over, for example, neural networks is the availability of uncertainty estimates which can be
used to make the observations more selective. In the present case, an observation (i.e., evaluation of the true energy and force)
was made for all intermediate images in each GP iteration. Alternatively, an observation may only be made for the image for which
there is greatest uncertainty. This could target the calculations better and thereby reduce the total number of energy and force evaluations
needed to converge to the true MEP.

While the whole path has to be converged well enough to provide an accurate estimate of the tangent, the part of the path
that is most important for practical purposes is the region around the first order saddle point.  In most cases, the MEP is needed mainly
to find the highest energy point along the path, i.e., the first order saddle point on the energy surface that is required for evaluating the 
transition rate within harmonic transition state theory. 
The algorithm can be refined to take this into account by, for example, applying the climbing-image NEB
\cite{ClimbingImage}, where one of the images is driven to the maximum energy along the path, and at the same time the tolerance for the convergence of other images can be increased.

In a typical case, the goal is to evaluate the transition rate using harmonic transition state theory. There, the second derivative matrix,
the Hessian matrix, and the frequency of vibrational modes need to be evaluated at the end points as well as at the (highest) first order 
saddle point.  While the saddle point is not known until the MEP calculation has been carried out, the minima are, and the second 
derivative matrices at those points might as well be calculated right from the start. This would provide additional information that could be
fed into the GP regression so as to improve the accuracy of the approximate energy surface right from first GP iteration. It remains an
interesting challenge to extend the GP regression approach to include in some way such information on the second derivatives. 

The test problems studied here are quite simple, and it will be important to test the method on more complex systems to 
fully assess its utility and to develop it further. One issue that can arise is that more than one MEP connects the two endpoint minima.
Then, some kind of sampling of MEPs needs to be carried out \cite{Maras16}. Also, some energy surfaces have multiple local minima
and highly curved MEPs, which can lead to convergence problems unless a large number of images are included in the calculation. 
The scaling of the GP regression approach to such more challenging problems needs to be tested. There will, however, clearly be a
large set of important problems, such as calculations of catalytic processes which often involve rather small molecules adsorbed on
surfaces, where the complexity is quite similar to the heptamer island test problem studied here and where the GP regression is 
clearly going to offer a significant reduction in computational effort.

At low enough temperature, quantum mechanical tunneling becomes the dominant transition mechanism, and the task is then to 
find the minimum action path \cite{Jonsson11,Mills97,Mills98}. 
Calculations of tunneling paths require exploring the energy surface over a wider region than a calculation of MEPs,
and here again the GP regression approach can lead to a significant reduction in computational effort, even more 
than for MEP calculations since each iteration necessarily involves more observations and thereby more input for the modelling of the
energy surface.

The discussion has focused here on atomic rearrangements, but it will, furthermore, be interesting to apply the GP regression approach 
to magnetic transitions where the evaluation of the magnetic properties of the system is carried out using computationally intensive 
{\it ab initio} or density functional theory calculations. There, the relevant degrees of freedom are the angles defining the orientation of
the magnetic vectors, and the task is again to find MEPs on the energy surface with respect to those
angles \cite{bessarab_13a,bessarab_14,bessarab_14b}.



\section*{Acknowledgements}
HJ would like to thank Prof. Andrew Peterson at Brown University for helpful discussions.
This work was supported by the Academy of Finland (FiDiPro program grant no. 263294) and by the Icelandic Research Fund.



\end{document}